%% file: bormio08.tex
\documentclass[aps,rmp,showpacs,showkeys,twocolumn,eqsecnum]{revtex4}

\usepackage{amsmath,amssymb,amsbsy}
\usepackage{graphicx}
\usepackage[usenames]{color}
\usepackage{slashed}
\usepackage{slashed,amsmath}

\newcommand{\del}[1]{}
\newcommand{\red}[1]{#1}

\newcommand{\fig}[1]{fig.~#1}
\newcommand{\figref}[1]{fig.~\ref{#1}}

\newcommand{\textfrac}[2]{#1/#2}
\newcommand{\comment}[1]{}

\newcommand{\eb}{\begin{eqnarray}}
\newcommand{\ee}{\end{eqnarray}}
\newcommand{\ebNN}{\begin{eqnarray*}}
\newcommand{\eeNN}{\end{eqnarray*}}

\newcommand{\etal}{\textit{et al.} }
\newcommand{\etalEnde}{\textit{et al.}}

\newcommand{\tinyMM}[1]{\text{\begin{tiny}#1\end{tiny}}}

\newcommand{\g}[1]{\gamma^{#1}}

\newcommand{\tw}[1]{{#1}}
\newcommand{\pr}[1]{{#1}\,'}
\newcommand{\ppr}[1]{p^{\prime \, #1}}

\newcommand{\BG}{\text{\begin{tiny}BG\end{tiny}}}

\newcommand{\MeV}{\; \mathrm{MeV} }
\newcommand{\GeV}{\; \mathrm{GeV} }

\newcommand{\ePunkt}{\hspace{0.3cm}.}
\newcommand{\eKomma}{\hspace{0.3cm},}

\newcommand{\M}{\mathcal{M}}

\newcommand{\twopi}{\left( 2\pi \right) }

\newcommand{\eq}[1]{eq.~#1}

\allowdisplaybreaks
\begin{document}

\title{Inclusive results for electron and photon scattering in the resonance region} 

\author{O. Buss}
\email{oliver.buss@theo.physik.uni-giessen.de}
\homepage{http://theorie.physik.uni-giessen.de/~oliver}
\thanks{Work supported by DFG.}
\author{T. Leitner}
\author{U. Mosel}
\affiliation{Institut f\"ur Theoretische Physik, Universit\"at Giessen, Germany}

\author{L. Alvarez-Ruso}
\affiliation{Departamento de F\'{\i}sica Te\'orica and IFIC,
Universidad de Valencia - CSIC, Spain}

\date{\today}
\include{abstract}

\maketitle

\section{Introduction and Motivation}\label{intro}
\input{IntroMotivation}

\input{elementary}

\input{electron}

\section{Summary}
\input{summary}

\begin{acknowledgments}
The authors thank Lothar Tiator for making a compilation of the MAID form factors available to us and for his kind support. We are grateful to all members of the GiBUU group. This work was supported by Deutsche Forschungsgemeinschaft (DFG). 
\end{acknowledgments}

\bibliography{literatur}

\end{document}

%% file: abstract.tex
\begin{abstract}
We present a model for electron scattering off nuclei and photon absorption in the resonance energy region ($W \leqq 2\GeV$). The elementary $\gamma$/$\gamma^\star$ N-vertex is described using in-medium kinematics and up-to-date form factors for QE-scattering, $\pi$-production form-factors and resonance helicity amplitudes of the MAID analysis. We find good agreement with inclusive data on electron scattering off Oxygen. For photon absorption in Carbon we find a large impact of the momentum dependent mean-field acting on initial- and final-state baryons. 
\end{abstract}

%% file: IntroMotivation.tex
A wealth of information on nucleon structure functions and resonance properties  has been gathered in the scattering of electrons and photons off nucleons at intermediate energies ($\sim 0.5-2 \GeV$) (cf.~, e.g., the reviews~\cite{Foster:1983kn,Burkert:2005ak}). Scattering electrons and photons on nucleons being embedded in nuclei, it is even possible to probe the change of these properties due to a strongly-interacting environment. In the medium, resonances and nucleons acquire an additional complex self energy due to rescattering effects and correlations which lead to modified spectral functions. By direct comparison to the elementary scattering case, one tries to deduce such in-medium modifications. 
Within the last decades, several experimental groups (cf. ref.~\cite{Benhar:2006wy} for a review on electron scattering and  \cite{Frommhold:1994zz,Bianchi:1995vb,Muccifora:1998ct,Cetina:2000rw} for the photon absorption experiments) have measured the inclusive cross section for electron and photon off nuclear targets to a high precision.  Additionally, major experimental efforts are being undertaken to shed light on the in-medium properties of hadrons studying exclusive channels of photon and electron nucleus scattering, see e.g.~\cite{Messch,Trnka:2005ey,:2007mga}.
 
In this work we aim at an improvement of the description of inclusive and exclusive  reactions through a better description of in-medium physics. Here we focus on inclusive cross sections which are independent on the final state interaction. These final-state interactions can, e.g., be treated applying the Giessen Boltzmann-Uehling-Uhlenbeck (GiBUU) transport model~\cite{GiBUUWebpage,Buss:2006vh,Buss:2006yk}, which has been developed over the last 20 years to describe heavy ion collisions, photon-, electron-, pion- and neutrino-induced reactions within one unified transport framework~\footnote{Note that the GiBUU source code is now freely available for public download~\cite{GiBUUWebpage}.}. Since our model for the inclusive process uses the very same inputs as the transport model - in particular the same momentum distributions for the nucleons, same potentials and spectral functions - the extraction of exclusive cross sections will be a straightforward task.

First results for electron scattering off nuclei for beam energies between $700\MeV$ and $1500\MeV$ and virtualities of $Q^2\leq 0.632 \GeV^2$ have already been presented in \cite{Buss:2007sa,Buss:2007ar}. Within the same framework we now also address the absorption of real photons in nuclei. Additionally, we refine the model presented in \cite{Buss:2007sa,Buss:2007ar} by including more baryon resonances in the description. Employing impulse approximation, we calculate inclusive cross sections with full in-medium kinematics and realistic form factors for quasi-elastic scattering, pion production and resonance excitation.

As already stressed in \cite{Buss:2007ar}, there are three key issues in the theoretical understanding of photon- or electron-scattering off nuclei. First and foremost, one needs to model effectively the nuclear ground state and take into account the modification of the elementary $\gamma$/$\gamma^\star$-nucleon vertex within the nuclear medium. Furthermore, the study of exclusive channels such as pion production or nucleon knockout demands for a proper description of final state interactions of the produced particles with the nuclear medium, which  can ideally be addressed within our GiBUU framework. 

In~\cite{Buss:2007ar}, we outlined the importance of the electron scattering information on the modeling of the neutrino nucleus interaction. Studying this interaction one aims at an understanding of, e.g., axial form factors and the strange quark content of the nucleon. However, the primary aim of investigating the $\nu A$ process is related to the interpretation of present-day neutrino oscillation experiments. A quantitative understanding of the influence of nuclear effects on that cross section is crucial since most of the experiments use nuclei as targets~\cite{nuint}. In this respect, the description of electron induced processes  serves as a benchmark for the neutrino induced reactions, which were analyzed by Leitner \etal also within the GiBUU framework~\cite{Leitner:2006sp,Leitner:2006ww}. 

This article is structured in the following way. First we introduce our model for the interaction of real and virtual photons with nucleons embedded in a nucleus. Next, we discuss two distinct applications: electron scattering off complex nuclei and photon absorption.

%% file: elementary.tex
\section{Elementary input}
The resonance excitations play a dominant role in the spectrum of low-energy $\gamma^\star$N scattering (for a comprehensive review see, e.g., \cite{boffi}). Applying a partial wave analysis to the world data on photon and electron scattering, e.g., the MAID group \cite{MAIDWebsite,Drechsel:1992pn,Drechsel:1998hk} provides information about resonance properties. An important result of such an analysis are the so-called helicity amplitudes, which can be used to fix the resonance form factors \cite{AlvarezRuso:1997jr,AlvarezRuso:2003gj,Lalakulich:2006sw}. We enhanced the model of \cite{Buss:2007ar} by including, besides the $\Delta$ resonance, additional resonance states, in particular all states which are also included in the MAID analysis (cf. table \ref{table:resPhoto} for details). 
\begin{table}
\centering
\begin{tabular}{lccc}
\noalign{\vspace{-8pt}}
\toprule
name            &  spin    & isospin   & parity     \\ 
\hline
P$_{11}$(1440)  &  1/2 & 1/2 & +     \\
D$_{13}$(1520)  & 3/2 & 1/2  & - \\
S$_{11}$(1535)  & 1/2 & 1/2  & - \\
S$_{31}$(1620)  & 1/2 & 3/2 & - \\
S$_{11}$(1650)  & 1/2 & 1/2 & - \\
D$_{15}$(1675)  & 5/2 & 1/2 & - \\
F$_{15}$(1680)  & 5/2 & 1/2 & + \\
D$_{33}$(1700)  & 3/2 & 3/2 & - \\
P$_{13}$(1720)  & 3/2 & 1/2 & + \\
F$_{35}$(1905)  & 5/2 & 3/2 & + \\
P$_{31}$(1910)  & 1/2 & 3/2 & + \\
F$_{37}$(1950)  & 7/2 & 3/2 & + \\
\toprule
\end{tabular}
\caption{Resonances being included in our model.}
\label{table:resPhoto}
\end{table}
For the nucleon resonances with spin $S=1/2$ we use a notation similar to Devenish \etalEnde~\cite{PhysRevD.14.3063}. The hadronic current is defined by
\eb
j_{1/2}^\nu =  \overline{\phi}(\pr{p},s_f) J_{\mbox{\begin{tiny}R,1/2\end{tiny}}}^{\nu} u(\tw{p},s_i) \eKomma
\ee
where $u$ is the nucleon spinor and $\phi$ is the resonance spinor. So $p$ and $\pr{p}$ denote the momentum of incoming nucleon and outgoing resonance, $s_i$ and $s_f$ denote their spins. The vertex operator is parametrized for positive-parity resonances by
\eb
(J^{+}_{\mbox{\begin{tiny}R,1/2\end{tiny}}})^\nu=\frac{g_1}{(2m_n)^2}(Q^2 \gamma^\nu+\slashed{q} q^\nu)+\frac{g_2}{2m_n} i \sigma^{\nu \rho}q_\rho
\ee
and for negative-parity ones by
\eb
(J^{-}_{\mbox{\begin{tiny}R,1/2\end{tiny}}})^\nu=(J^{+}_{\mbox{\begin{tiny}R,1/2\end{tiny}}})^\nu \gamma_5 \ePunkt
\ee
The constant $m_n=0.938 \GeV$ denotes the isospin averaged nucleon mass. We assume that the form factors $g_1$ and $g_2$ depend solely on $Q^2$ and can be fixed using the MAID helicity amplitudes. The hadronic tensor is given by
\begin{eqnarray}
&&H_{\mbox{\begin{tiny}R,1/2\end{tiny}}}^{\mu\nu}
=\frac{1}{2}\sum_{s_i,s_f}
\overline{\phi}(\pr{p},s_f) J_{\mbox{\begin{tiny}R,1/2\end{tiny}}}^{\mu} u(\tw{p},s_i) \nonumber\\
&& \times \left(\overline{\phi}(\pr{p},s_f) J_{\mbox{\begin{tiny}R,1/2\end{tiny}}}^{\nu} u(\tw{p},s_i)\right)^\star \\
&&=\frac{1}{2} Tr \left[ 
\frac{\pr{\slashed{p}}+\pr{m}}{2\pr{m}} J_{\mbox{\begin{tiny}R,1/2\end{tiny}}}^{\mu} \frac{\tw{\slashed{p}}+\tw{m}}{2\tw{m}}\g{0} (J_{\mbox{\begin{tiny}R,1/2\end{tiny}}}^{\nu})^\dagger \g{0} \right]
\end{eqnarray}
with $\pr{m}=\sqrt{\pr{p}\cdot \pr{p}}$ and $m=\sqrt{p\cdot p}$ being the incoming and outgoing baryon masses.

The spin $3/2$ resonances can be described using Rarita-Schwinger spinors $\psi_\alpha$. Within this frame\-work, the hadro\-nic current is given by
\eb
j_{1/2}^\nu =  \overline{\psi}_\alpha(\pr{p},s_f) J_{\mbox{\begin{tiny}R,3/2\end{tiny}}}^{\alpha\nu} u(\tw{p},s_i) \eKomma
\ee
where $u$ is the nucleon spinor and $\psi_\alpha$ is the resonance Rarita-Schwinger spinor. The vertex operator for negative-parity resonances can be parametrized by
\begin{eqnarray}
(J^{-}_{\mbox{\begin{tiny}R,3/2\end{tiny}}})^{\alpha \nu}&=&
g^{\alpha \nu}\left(\frac{C_3}{m_n}\slashed{q}+\frac{C_4}{m_n^2}\ppr{}\cdot q+\frac{C_5}{m_n^2} p\cdot q \right)\nonumber\\
 &&
-q^\alpha \left( \frac{C_3}{m_n} \gamma^\nu+\frac{C_4}{m_n^2}\ppr{\nu}+\frac{C_5}{m_n^2} p^\nu
\right)
\end{eqnarray}
and for positive-parity ones by
\eb
(J^{+}_{\mbox{\begin{tiny}R,3/2\end{tiny}}})^\nu=(J^{-}_{\mbox{\begin{tiny}R,3/2\end{tiny}}})^\nu \gamma_5 \ePunkt
\ee
Again, we assume that the form factors $C_3$, $C_4$ and $C_5$ depend solely on $Q^2$; these form factors are again fixed using helicity amplitudes. The hadronic tensor is given by \begin{eqnarray}
&& H_{\mbox{\begin{tiny}R,3/2\end{tiny}}}^{\mu\nu}
=\frac{1}{2}\sum_{s_i,s_f}
\overline{\psi}_\alpha(\pr{p},s_f) J_{\mbox{\begin{tiny}R,3/2\end{tiny}}}^{\alpha \mu} u(\tw{p},s_i) \nonumber\\
&&       \times \left(\overline{\psi}_\beta(\pr{p},s_f) J_{\mbox{\begin{tiny}R,3/2\end{tiny}}}^{\beta \nu} u(\tw{p},s_i)\right)^\star \\
&&=\frac{1}{2} Tr \left[ 
\frac{\Lambda_{\beta\alpha}}{2\pr{m}} J_{\mbox{\begin{tiny}R,3/2\end{tiny}}}^{\alpha\mu} \frac{\tw{\slashed{p}}+\tw{m}}{2\tw{m}}\g{0} (J_{\mbox{\begin{tiny}R,3/2\end{tiny}}}^{\beta\nu})^\dagger \g{0}  \right] .
\end{eqnarray}
In the equation above, we used the identity for the spin 3/2 projector 
\eb
 \sum_{s_f} \psi_\beta(\pr{p},s_f)  \overline{\psi}_\alpha(\pr{p},s_f) = \frac{\Lambda_{\beta\alpha}}{2\pr{m}}
\ee
with\pagebreak[4]
\eb
\Lambda_{\beta\alpha}(p)&=&-(\pr{\slashed{p}} + \pr{m}) \left( g^{\beta\alpha}  -\frac{2}{3} \frac{ \ppr{\beta}\ppr{\alpha}}{(\pr{m})^2} \right. \nonumber\\
&& + \left. \frac{1}{3} \frac{\ppr{\beta}\gamma^\alpha - \ppr{\alpha}\gamma^\beta}{\pr{m}} - \frac{1}{3} \gamma^\beta \gamma^\alpha\right)\, .
\label{eq:spin3_2_projector}
\ee

The description of $5/2$ spinors within a Lagrange framework is highly complicated.  As a simplifying assumption, we will treat all particles with spin greater than $3/2$ within the  spin $3/2$ formalism.

Additionally to this improvement of the resonance description, we updated our elastic nucleon form factors to the ones presented by Bradford~\etalEnde~\cite{Bradford:2006yz}.

\subsection{Total cross sections}
For \emph{electron induced} events, the total cross section  contains quasi-elastic scattering, resonance production and direct pion production (cf. \cite{Buss:2007ar} for the implementation of quasi-elastic scattering and direct pion production). For the \emph{photon induced} events only  resonance excitation and pion production are relevant, because one may neglect the QE contribution since the outgoing nucleon has large off-shellness for $Q^2=0$.  In both cases one must be careful to avoid double counting since also the resonances contribute to pion production. Thus one must subtract these resonance contributions from the direct pion production cross section. We interpret this subtracted contribution as a single-$\pi$ background and denote it by $\sigma^{\BG}_\pi$. With this prerequisite, we define the total cross sections for the electron
\begin{eqnarray}
\frac{d\sigma_{e,\tinyMM{tot}}}{d\Omega_{\pr{l}}~ d|\vec{\pr{l}}|}&=&
\frac{d\sigma_{\tinyMM{QE}}}{d\Omega_{\pr{l}}~ d|\vec{\pr{l}}|}
+\frac{d\sigma^{\BG}_\pi}{d\Omega_{\pr{l}}~ d|\vec{\pr{l}}|} \\
&& +\sum_{R}\frac{d\sigma_\tinyMM{R}}{d\Omega_{\pr{l}}~ d|\vec{\pr{l}}|} \eKomma
\label{eq:sigTot_electron}
\end{eqnarray}
where $l$ and $\pr{l}$ are the momenta of incoming and outgoing electron, respectively. For the photon we get
\begin{eqnarray}
\sigma_{\gamma,\tinyMM{tot}}=\sigma^{\BG}_\pi+\sum_{R}\sigma_\tinyMM{R} \;\;\; .
\label{eq:sigTot_photon}
\end{eqnarray}
Note that the background terms $\sigma_\pi^\text{BG}$ effectively include interferences among the resonances. 

\begin{figure}
 \centering
 \includegraphics[width=0.45 \textwidth]{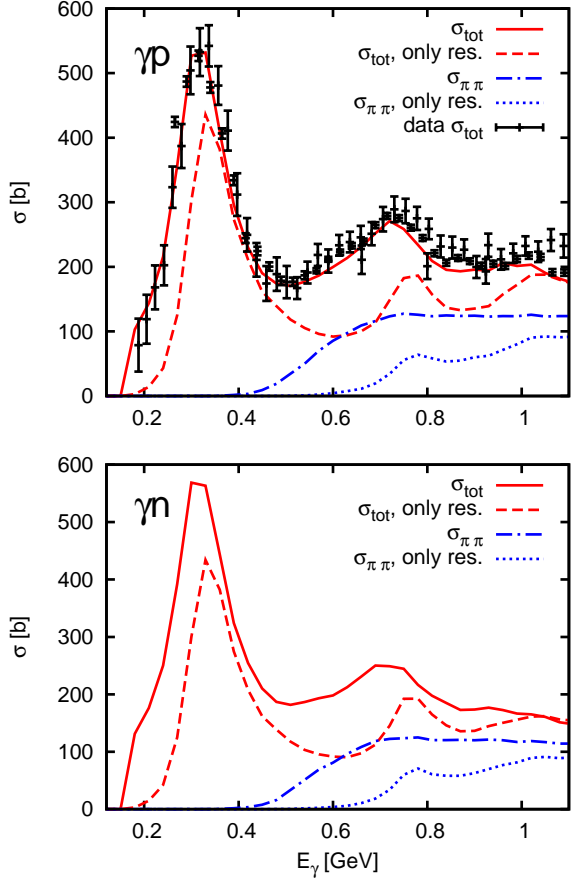}
\caption{Photon scattering off proton (upper panel) and neutron (lower panel) as a function of the incoming photon energy including a $2\pi$ background. The solid line shows the result for $\sigma_\text{tot}$ for the full model, the dashed line shows the result without single-$\pi$ and  $\pi\pi$ background contributions. The contribution of the resonances to double-$\pi$ production is shown by the dotted line, the full $2\pi$ contribution is shown by the dashed-dotted line. The compilation of data is taken from PDG~\cite{PDBook}.}
\label{fig:photonEle}
\end{figure}

The cross section for photon scattering off the proton and neutron are shown in \figref{fig:photonEle}. Evidently, the resonance model alone describes the cross sections only qualitatively. Additionally to the single-$\pi$ background discussed in \cite{Buss:2007ar}, we included for the photon induced reactions also a $\pi\pi$ production background. Given parametrizations $\sigma_{\pi\pi}^\text{data}$ of the elementary data, we can define the $2\pi$ background by subtracting the resonance contributions from the parametrization 
\[
\sigma_{\pi\pi}^{BG}=\sigma_{\pi\pi}^\text{data}-\sigma_{\pi\pi}^\text{Res} \; .
\]
This subtraction must be done independently for all possible charge channels: $\pi^+\pi^-$,$\pi^0\pi^0$ and $\pi^0\pi^\pm$. In our model, double pion production via resonances can only occur via the four channels
\ebNN
\begin{array}{lllllll}
\gamma N&\to& R &\to& \pi \Delta 	&\to &2\pi N \eKomma\\
\gamma N&\to& R &\to& \pi P_{11}(1440) 	&\to &2\pi N \eKomma\\
\gamma N&\to& R &\to& \rho N 		&\to &2\pi N \eKomma\\
\gamma N&\to& R &\to& \sigma N 		&\to &2\pi N  \ePunkt
\end{array}
\eeNN
Given $\sigma_{\gamma N\to R}$, the evaluation of $\sigma_{\gamma N\to R\to N\pi\pi}$ involves  a weighting with the partial decay widths (for details cf. ref.~\cite{Buss:2006vh} and references therein) into the channels listed above and the relevant isospin Clebsch-Gordan factors. The distribution of the final state momenta of the $\pi\pi$ background events are assumed to follow a phase space distribution. The resulting $2\pi$ contributions are shown in \fig{\ref{fig:photonEle}} for both proton and neutron targets. Obviously, the total cross section is now very well described and the resonance contribution to $\pi\pi$ is in fact small as compared to the total $\pi\pi$ production cross section, especially at energies below 600 MeV. 

\begin{figure}
 \centering \includegraphics[width=0.45 \textwidth]{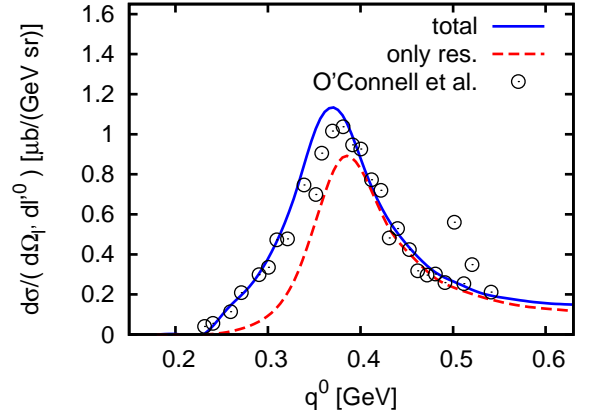}
\caption{Electron scattering off a proton at an incoming electron energy of 730 MeV and an electron scattering angle of $37.1^\circ$. The plot shows a comparison of our elementary cross section with and without background to the experimental data obtained by O'Connell et al.~\cite{OConnell1984} in lab coordinates as a function of the virtual photon energy $q^0$. The statistical errors of the data points are negligible, therefore they have been omitted.}
\label{fig:resBG730}
\end{figure}
Furthermore, we show in \figref{fig:resBG730} the single-$\pi$ production cross section for electron induced reactions with and without background contributions. Also here, the background has a considerable impact and must be taken into account. However, we can not introduce a sensible $2\pi$ background due to a lack of such data, and thus we consider only a single-$\pi$ background. This restricts our model to the region not far above the $\pi \pi$ threshold.

\subsection{Medium modifications}
As detailed in \cite{Buss:2007ar}, we consider, besides momentum-dependent mean fields for the baryons, also the in-medium broadening of the resonances based on the GiBUU collision rates. The quasi-elastic and resonance cross sections are evaluated with full in-medium kinematics. For the backgrounds we assume that the in-medium background is given by
\eb
\sigma_\text{bg,in-med}(s)=\sigma_\text{bg,free}(s_\text{free})
\ee
where $s_\text{free}$ is the Mandelstam $s$ corrected for the in-medium potentials:
\eb
s_\text{free}=(q+p_\text{free})^2 \eKomma
\label{eq:sfree}
\ee
where 
\[
p_\text{free}=(\sqrt{m_N^2+\vec{p}\,^2},\vec{p})\ePunkt 
\]

\subsection{Nuclear cross sections} 
Assuming a small wavelength of the incoming photon, we will treat the whole problem in the impulse approximation (IA). Within IA, the nuclear current operator is assumed to be a sum of one-body currents $J_A\to \sum_i J_i^\mu$.
Furthermore, we assume that the nucleus is constructed out of a sum of single particle plane-wave states. Note, that the dispersion relations of the plane waves are modified due to the potentials. In Bjorken-Drell notation~\cite{BjorkenDrell}, each plane wave state $\left|k,s\right\rangle$ ($s$ denotes spin) has the normalization $1/\sqrt{2m(k)/(2E_k)}$. So the nuclear wave function is given by
\eb
\Psi(r)&=&\sum_{s=\pm \frac{1}{2}} \left(  \int^{k_F(\text{protons} )} \frac{d^3 k}{\twopi^3}  \; .
\sqrt{\frac{2m(k)}{2E_k}} \left|k,s\right\rangle \right. \nonumber \\
&&+\left.  \int^{k_F(\text{neutrons})} \frac{d^3 k}{\twopi^3}  \; .
\sqrt{\frac{2m(k)}{2E_k}} \left|k,s\right\rangle \right) \; .
\ee
The latter wave function is normalized to the number of nucleons $A$
\[
 \int d^3r \Psi^\dagger(r)\Psi(r)=A \; .
\]
\subsubsection{Electron scattering.}
Within the above approximations, the matrix element for electron scattering off a nucleus with mass number $A$ is given by a contraction of the lepton tensor with the sum of all hadronic currents generated by the $A$ nucleons within the nucleus. In formulas, we get (cf. also \cite[in particular \eq{1-3}]{De_Forest:1983vc})
\[
|\M_A|^2= L_{\mu\nu} \int d^3 r  H^{\mu\nu}_{A}(r) 
\]
where 
\ebNN
&&H^{\mu\nu}_{A}(r)
=\frac{1}{2}\sum_{s,s^\prime,\ldots}\sum_{\alpha=p,n}\int^{p^\alpha_f(r)} \frac{d^3 p}{\twopi^3}  \sqrt{\frac{2m(p)}{2E_p}}\\
&&\times \sqrt{\frac{2m(p^\prime)}{2E_{p^\prime}}} \times \ldots \times 
\left\langle p^\prime,s^\prime;\ldots  \right|   J^\mu       \left|p,s\right\rangle \\
&& \times  \sqrt{\frac{2m(p)}{2E_p}}\sqrt{\frac{2m(p^\prime)}{2E_{p^\prime}}} \times \ldots \times 
\left(\left\langle p^\prime,s^\prime;\ldots  \right|   J^\nu       \left|p,s \right\rangle\right)^\star\\
&&=\sum_{\alpha=p,n} \int^{p_f^\alpha(r)} \frac{d^3 p}{\twopi^3}  \frac{2m(p)}{2E_p}  \frac{2m(p^\prime)}{2E_{p^\prime}} \times \ldots \\
&& \times  
\underbrace{\frac{1}{2}\sum_{s,s^\prime,\ldots}\left\langle p^\prime,s^\prime;\ldots \right|   J^\mu       \left|p,s\right\rangle 
\left(\left\langle p^\prime,s^\prime;\ldots \right|   J^\nu       \left|p,s \right\rangle\right)^\star}_{{H}^{\mu\nu}} \; .
\eeNN
Note that the dots "$\ldots$" in the brackets should just remind us that there could be more than one final state particle. Obviously, the previous equation is just an integral over single particle hadronic currents ${H}^{\mu\nu}$. Thus the nuclear cross section is given by
\eb
&&d\sigma_A=\sum_{\alpha=p,n} \frac{1}{v_\text{rel}}\frac{2m_e}{2l_0} \frac{2m_e}{2l^\prime_0} L_{\mu\nu}  \int d^3r \int^{p_f^\alpha(r)} \frac{d^3 p}{\twopi^3} \nonumber\\  && \int d\Phi_f   \frac{2m(p)}{2E_p}  \frac{2m(p^\prime)}{2E_{p^\prime}}  \times \ldots
\times {H}^{\mu\nu}(r,p,\Phi_f) \nonumber\\
&&\times \twopi^4 \delta^4(p+l-l^\prime-p^\prime-\ldots)  \nonumber       P_{PB}(\vec{r},\Phi_f)  
\ee
where $v_\text{rel} \simeq 1$ is the relative velocity of nucleus and electron. The variable $\Phi_f$ denotes the phase space of all the final state particles, which includes also $p^\prime$ and $l^\prime$. The function $P_{PB}(\vec{r},\Phi_f)$ includes Pauli blocking of the final state fermions. Bose enhancement of the pions can be safely neglected. Comparing to the single particle cross sections $\sigma_N$, one finds that
\eb
&& d\sigma_A= \sum_{\alpha=p,n}\int d^3r \int^{p_f^\alpha(r)} \frac{d^3 p}{\twopi^3} \;\frac{|v_N-v_e|}{v_\text{rel}} d\sigma_N \nonumber\\
&&\approx
\sum_{\alpha=p,n}\int d^3r \int^{p_f^\alpha(r)} \frac{d^3 p}{\twopi^3} \;\frac{1}{v_\text{rel}}\frac{l_\mu p^\mu}{l_0 p_0} d\sigma_N \eKomma
\label{eq:elektronKernCrossSection}
\ee
where the factor $\textfrac{l_\mu p^\mu}{(l_0 p_0)}$ is just a flux correction due to the finite velocity of the target nucleons.
The variable $v_N$ denotes the non-zero nucleon velocity, which is due to Fermi motion, and $v_e$ is the electron velocity. Note, that the Pauli blocking factor has been included in the single particle cross section. 

\subsubsection{Photon scattering.} For photon scattering we get an analogous result \eb
&&d\sigma_A= \sum_{\alpha=p,n}\frac{1}{v_\text{rel}}\frac{1}{2q_0}    g_{\mu\nu}  \int d^3r \int^{p_f^\alpha(r)} 
\frac{2m(p)}{2E_p}  \frac{2m(p^\prime)}{2E_{p^\prime}} \nonumber\\
&& \times \ldots\times {H}^{\mu\nu}(r,p,\Phi_f) \frac{d^3 p}{\twopi^3}                \;            P_{PB}(\vec{r},\Phi_f) d\Phi_f \eKomma
\ee
\eb
&&d\sigma_A
=\sum_{\alpha=p,n}\int d^3r \int^{p_f^\alpha(r)}\;\frac{d^3 p}{\twopi^3} \frac{|v_N-v_\gamma|}{v_\text{rel}} d\sigma_N \nonumber\\
&&=
\sum_{\alpha=p,n}\int d^3r \int^{p_f^\alpha(r)}\;\frac{d^3 p}{\twopi^3} \frac{1}{v_\text{rel}}\frac{q_\mu p^\mu}{q_0 p_0} d\sigma_N 
\label{eq:photonKernCrossSection}
\ee
with $v_\gamma$ being the photon velocity vector. Again, we included the Pauli blocking probability in the single particle cross section.

%% file: electron.tex
\section{Inclusive electron scattering off complex nuclei}
There is a considerable amount of theoretical work aiming at a good description of the inclusive electron cross section; cf. the recent review given by Benhar \etalEnde~\cite{Benhar:2006wy} for an overview. Benhar and collaborators~\cite{Benhar:2005dj} employ the impulse approximation with realistic spectral functions obtained from electron-induced proton knockout data and theoretical calculations based on nuclear many body theory (NMBT). With this model, they achieve impressive agreement in the quasi-elastic (QE) peak region; however, they underestimate the data in the $\Delta$ region. 
In~\cite{Benhar:2006qv,Nakamura:2007pj} they improved on this and a good description of the data also in the single-pion production region could be reached. 
Also, Szczerbinska \etalEnde~\cite{Szczerbinska:2006wk} use Benhar's spectral functions~\cite{Benhar:2005dj} for the QE contribution, but in the $\Delta$ region they apply the dynamical Sato-Lee model developed to describe photo- and electron-induced pion-production off the nucleon. Recently, this model has been extended to weak-interaction processes~\cite{Sato:2000jf,Sato:2003rq}. 
A different approach, which also yields a combined investigation of neutrino and electron interactions, makes use of the superscaling properties of the electron scattering data (cf.~\cite{Amaro:2004bs} and references therein). 
There the authors extract the scaling function from inclusive electron-nucleus scattering data and use this to predict the neutrino-nucleus cross sections. 
More work has been done in the QE region. In particular the model by Gil \etalEnde~\cite{Gil:1997bm} yielded a  particular successful description of the dip region in between the QE and $\Delta$ peak.

\subsection{Results}
\begin{figure}
   \centering
  \includegraphics[width=0.5\textwidth]{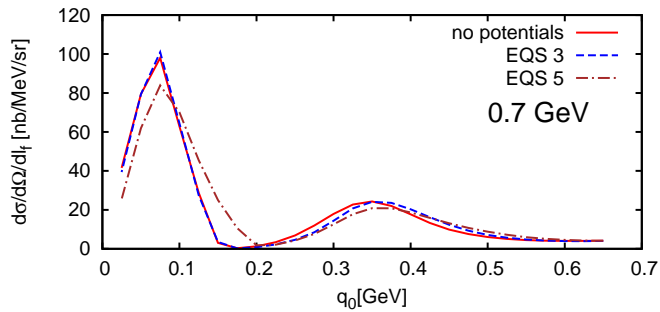} 
\caption{The inclusive electron cross section $d\sigma / (d\Omega dl_f)$ on $^{16}$O  as a function of the energy transfer $q_0=k_0-k'_0$ for a beam energy of 0.7 GeV and a scattering angle of $\theta_{l_f}=32^\circ$. The graph shows the results for different nucleon potentials: no potential (solid line), momentum-independent potential EQS 3 (dashed line) and momentum-dependent potential EQS 5 (dashed-dotted line). The calculations do not include in-medium changes of the widths.}
\label{fig:electron_EQS}  
\end{figure}

To start the discussion, we show in \fig{\ref{fig:electron_EQS}} our results for the inclusive reaction $^{16}_8O\left(e^-,e^-\right)X$ for a beam energy of 700 MeV and different nucleon mean fields, in-medium changes to the width have been neglected. The solid curve denotes the result without potentials, including only Fermi motion and Pauli blocking. Including a momentum-independent potential (dashed curve) does not change anything at the QE peak ($q_0=0.-0.15 \GeV$). However, the single-pion region ($q_0\gtrsim 0.2 \GeV$) is modified. This is due to the fact that the $\Delta$ is less strongly bound than the nucleon. Therefore, more energy must be transferred by the photon such that this binding effect is compensated. When the momentum dependent mean field is included (dashed-dotted curve), then the faster (on average) final state nucleons experience a shallower potential than the initial state nucleons. Also the resonance potential gets shallower towards higher momentum. Therefore, even more energy must be transferred by the photon. Hence, the QE peak is broadened towards a higher energy transfer $\omega$ and also the single pion spectrum is slightly shifted towards higher energies and broadened. 
A similar result has also been obtained within the Walecka model \cite{Rosenfelder:1980nd}.  There the nucleon mass becomes an effective mass $m^\star(\vec{r})$ and the energy of the nucleon is given by 
$
E_{\vec{p}}=\sqrt{\vec{p}\,^2+m^\star(\vec{r})} 
$
, which can be re-written as
$
E_{\vec{p}}=\sqrt{\vec{p}\,^2+m} +V(\vec{r},\vec{p})
$ 
with the momentum-dependent potential
\[
V(\vec{r},\vec{p})=\sqrt{\vec{p}\,^2+m^\star(\vec{r})}-\sqrt{\vec{p}\,^2+m} \; .
\]
E.g. for small momenta ($|\vec{p}|\ll m^\star$, $|\vec{p}|\ll m$) one obtains
a simple harmonic dependence of the potential on the momentum: 
\[
V(\vec{r},\vec{p})\approx p^2 \frac{m-m^\star(\vec{r})}{2mm^\star(\vec{r})}+m^\star(\vec{r})-m \ePunkt
\]
Rosenfelder~\cite{Rosenfelder:1980nd} shows that the value of $m^\star$ can then be used to fit the QE peak. We emphasize however, that we do not fit our potential to the electron data but the potential has been fixed by nuclear matter properties and the momentum dependence of  nucleon-nucleus scattering cross sections~\cite{Teis:1996kx}.

\begin{figure}
   \centering
 \includegraphics[width=0.45\textwidth]{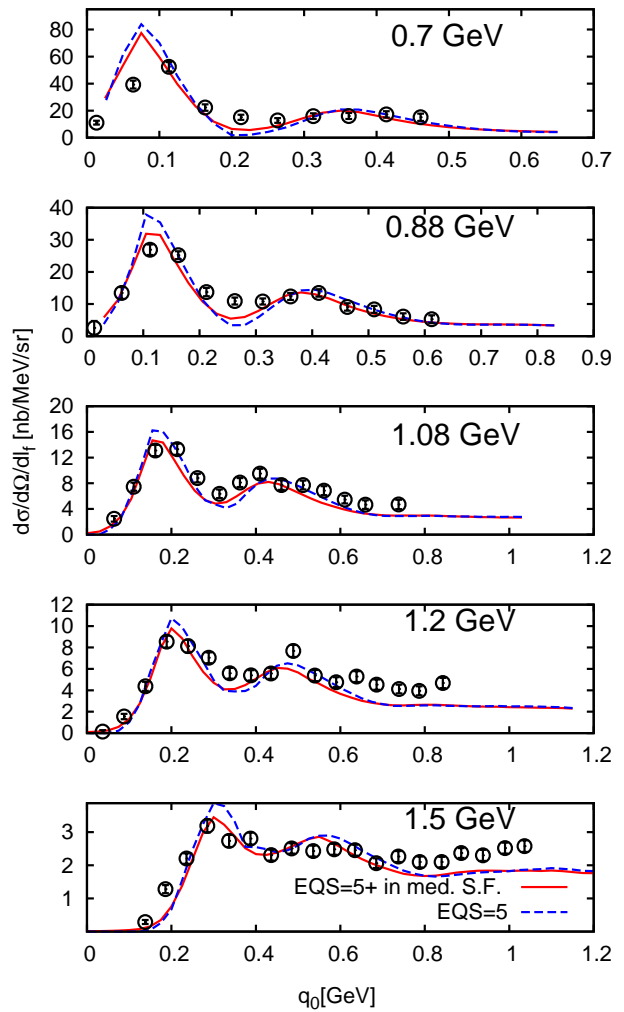} 
\caption{The inclusive electron cross section $d\sigma / ((d\Omega dl_f))$ on $^{16}$O  as a function of the energy transfer $q_0=k_0-k'_0$ at five distinct fixed electron energies (0.7, 0.88, 1.08, 1.2 and 1.5 GeV) and a scattering angle of $\theta_{l_f}=32^\circ$. The dashed line denotes our result, where we include all in-medium modifications besides collisional broadening.  The solid line denotes the full calculation, which includes in-medium changes of the width according to the GiBUU collision term. The data are taken from~\cite{Anghinolfi:1996vm,QEWebsite}.  }
\label{fig:electronAll}  
\end{figure}

Fig.~\ref{fig:electronAll} shows the comparison of our full calculation, which includes both a momentum dependent potential and collisional broadening of the baryons to the data measured by Anghinolfi \etalEnde~\cite{Anghinolfi:1996vm}. One observes that the agreement  with the data is considerably improved when the in-medium width is included. However, we also notice a short-coming of our model for the QE-region in the upper panel (beam energy=0.7 GeV). This problem could not yet been resolved. In the work of Kalok~\cite{DavidDiplom} it was shown that the inclusion of short-range-correlations in our model lead to a lowering of the peak. These short-range correlations imply a modified momentum distribution of the ground state nucleus. However, also with Kalok's improvements~\cite{DavidDiplom} no quantitative satisfactory result could be retrieved.  We obtain already at a slightly higher beam energy of 880 MeV (cf. \fig{\ref{fig:electronAll}}) very good agreement with the experimental data. The solid curve in \fig{\ref{fig:electronAll}} represent our full result which agrees well with the available data. Especially the dip-region in between QE-region and single-pion region is well reproduced. At very high beam energies, a lack of $\pi\pi$ strength leads to a too low result at large photon energies. Overall, the in-medium width leads to an improvement of the model.

\begin{figure}
\centering
\includegraphics[width=0.45\textwidth]{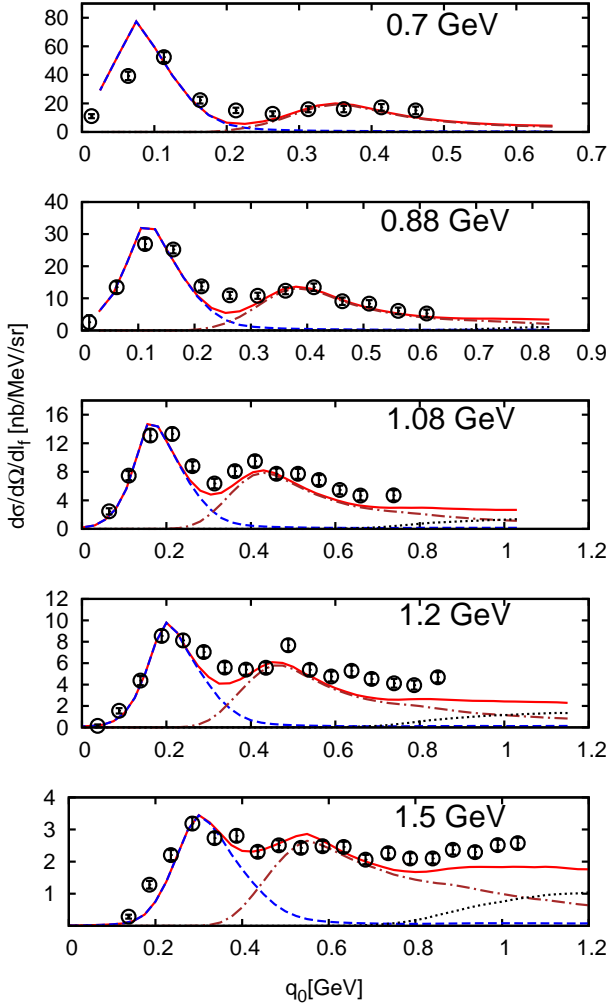} 
\caption{The inclusive electron cross section $d\sigma / ((d\Omega dl_f))$ on $^{16}$O  as a function of the energy transfer $q_0=k_0-k'_0$ at five distinct fixed electron energies (0.7, 0.88, 1.08, 1.2 and 1.5 GeV) and a scattering angle of $\theta_{l_f}=32^\circ$. The solid line denotes our full result, where we include all in-medium modifications and in particular in-medium changes of the width. The data are taken from~\cite{Anghinolfi:1996vm,QEWebsite}. The dashed line shows the quasi-elastic contribution, the dashed-dotted the single-$\pi$ and the dotted one the $2\pi$ contribution to the initial scattering process. This result does not yet include any FSI of the outgoing particles.}
\label{fig:electronAll_channels}  
\end{figure}

In \fig{\ref{fig:electronAll_channels}}, we show the contribution of the different mechanisms to the total electron-nucleus cross section which we calculated including all in-medium modifications and in particular the in-medium changes of the width. The dashed line shows the quasi-elastic contribution, the dashed-dotted the single-$\pi$ and the dotted one the $2\pi$ contribution to the initial scattering process. One observes that going from low to high beam energies, the importance of single-$\pi$ and $2\pi$ production mechanisms gradually increases, whereas at low energies the quasi-elastic contribution is dominating. Note that this result does not yet include any FSI of the outgoing particles and the classification into different channels is solely based on the initial vertex and not on the final-state multiplicities.

\begin{figure}
   \centering
 \includegraphics[width=0.5 \textwidth]{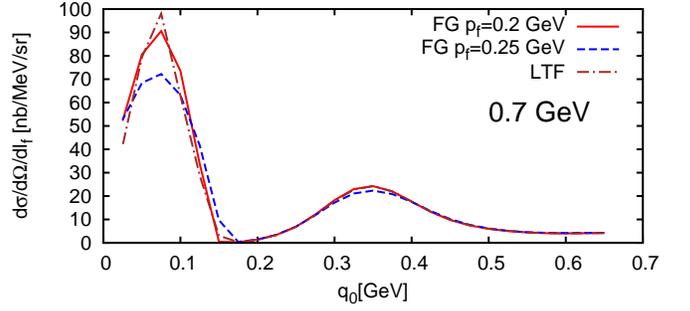} 
\caption{The inclusive electron cross section $d\sigma / (d\Omega dl_f)$ on $^{16}$O  as a function of the energy transfer $q_0=k_0-k'_0$ for a beam energy of 0.7 GeV and a scattering angle of $\theta_{l_f}=32^\circ$. The graph shows the results for various assumptions concerning the momentum distribution of the target nucleons: Fermi gas with Fermi momentum $p_f=0.2 \GeV$ (solid line), Fermi gas with Fermi momentum $p_f=0.25 \GeV$ (dashed line), momentum distribution according to local Thomas-Fermi (LTF) approximation (dashed-dotted line). The calculations do not include any potentials or in-medium changes of the widths.}
\label{fig:electron_FG}  
\end{figure}

To point out the importance of the momentum distribution of the initial nucleons, we show in \fig{\ref{fig:electron_FG}} the results for different input distributions. In this case we omit all in-medium modifications besides Pauli blocking and Fermi motion. The dashed-dotted curve represents the result with our standard momentum distribution according to the local-Thomas-Fermi (LTF) ansatz (cf. \cite{Buss:2007ar} for details). Distributing the nucleons according to a constant Fermi momentum of $0.2 \GeV$ yields already a slightly lower QE-peak value. With a further increase of the Fermi momentum to $0.25 \GeV$ we conceive a prominent drop of the QE-peak height and a broadening of the QE peak. It is therefore also clear, that one can fit the data for the QE-peak by a variation of the Fermi momentum. However, a Fermi momentum of $0.25 \GeV$ does not make any sense within our framework: on the basis of a proper density profile for the dilute Oxygen nucleus one can not get such a high average Fermi-momentum within LTF approximation.

We conclude, that the overall agreement to the data is improved by a calculation which in addition to a local Fermi-gas momentum-distribution also includes a momentum-dependent mean field and in-medium spectral functions for the nucleon. Especially at low energies, a proper treatment of the nucleon spectral function is important. The increase of the energy loss due to a momentum-dependent nucleon potential reshapes the QE peak considerably. An additional modification of the nucleon width leads to further broadening and decrease of the QE peak height. In the single-$\pi$ and $\Delta$ production region, we achieve a good description for all energies. The in-medium modifications improve the overall correspondence with the data. In the dip region, which is conventionally attributed to 2N excitations, the description is considerably improved due to the previously discussed broadening of the QE peak. At higher beam energies, the data are underestimated at high energy transfer $q_0$ due to the fact that $2\pi$-production channels have not yet been included. 

An impulse approximation calculation by Benhar \etalEnde~\cite{Benhar:2005dj} that uses nuclear many-body theory (NMBT) spectral functions yields in the QE-region a better result for 700 MeV beam energy. However, already at a slightly higher beam energy of 1080 MeV our model and the NMBT one yield equally good results for the QE peak. We thus conclude, that our simplified ansatz for the in-medium width and the inclusion of a proper potential incorporate the main features of the nucleon spectral function in the medium. 

\section{Total photon absorption cross sections for complex nuclei}
In the total photon absorption cross section on nucleons, one observes three major peaks, which are, according to our present knowledge, generated by several overlapping resonances. The most important ones are the $P_{33}(1232)$, $S_{11}(1535)$, $D_{13}(1520)$ and $F_{15}(1680)$ resonance states. To study the properties of those resonance states when embedded in nuclear matter, one has to investigate their  photon-induced excitation in nuclei. First experiments using a tagged high energy photon-beam ($E_\gamma=0.3-2.6 \GeV$), which offered sufficient energy to excite the second resonance region, were performed by the Yerevan group~\cite{Arakelian:1982ba,Ananikyan:1987}. Following up this pioneering work, the photon absorption of photons in nuclei was measured at the \textit{Mainz microton (MAMI) facility}~\cite{Frommhold:1992um,Frommhold:1994zz} with a beam energy of $E_\gamma=0.05 - 0.8\GeV$, with  higher energy of $E_\gamma=0.2 - 1.2 \GeV$ at the \textit{Adone storage ring facility} (Frascati, Italy) \cite{Bianchi:1992ze,Bianchi:1993nh,Bianchi:1993mb,Bianchi:1994ax,Bianchi:1995vb}, with even higher energy using the \textit{SAPHIR tagged photon beam} of $E_{\gamma}=0.5-2.67 \GeV$ at \textit{ELSA} (Bonn, Germany) \cite{Muccifora:1998ct}
and with $E_{\gamma}= 0.17- 3.84 \GeV$ at \textit{Hall B} of the \textit{Jefferson Laboratory} (Newport News, USA) \cite{Cetina:2000rw,Cetina:2002mz}.  In fact, some of the above experiments did not measure directly the photon absorption cross section but only the photo-fission cross-section \cite{Frommhold:1992um,Frommhold:1994zz,Bianchi:1993mb,Bianchi:1992ze,Cetina:2000rw,Cetina:2002mz}. Contradicting to earlier assumptions, it has been shown in \cite{Cetina:2000rw,Cetina:2002mz} that these two cross sections must not be identical. So we focus for our analysis on a comparison with the direct measurements of photon absorption as presented by Bianchi \etalEnde~\cite{Bianchi:1994ax,Bianchi:1995vb} and  Muccifora \etalEnde~\cite{Muccifora:1998ct}.

Let us briefly summarize the experimental findings. First, the $\Delta$ resonance region is also within the nucleus still exhibiting a peak like structure. However, one has observed a slight shift to higher energies and a broadening as compared to the vacuum structure. At higher energies, one has observed a more interesting and a somewhat unexpected effect. The experimental results show no structures in the second and third resonance region, often quoted as the disappearance of the resonances in the medium. 

In the last two decades, various theoretical attempts have been performed to explain these data. 
Kontratyuk \etalEnde~\cite{Kondratyuk:1993ah} set up a baryon resonance model including the collisional widths of the resonances as a free parameter. Then these parameters were fitted to the existing data and the collisional width was extracted. However, the extracted widths are extraordinary large (ca. $320 \MeV$ for the  $S_{11}(1535)$, $D_{13}(1520)$ and $F_{15}(1680)$ resonance) and the whole analysis offers several points for criticism (cf. pages 368-369 in \cite{Effenberger:1996im}).
The elaborated work of Carrasco \etalEnde~\cite{Carrasco:1989vq} showed in a microscopic model approach, that it is possible to describe the data in the $\Delta$ resonance region in a very satisfactory manner when including the $\Delta$ self-energy. This work emphasizes the importance of multi-body absorption channels.
Rapp \etalEnde~\cite{Rapp:1997ei} applied a vector-meson dominance (VMD) model to the problem. An energy-independent collisional broadening of the baryon resonances of $15\MeV$ for the $\Delta$, $250 \MeV$ for the $D_{13}$ and $50 \MeV$ for all other contributing resonances has been included by hand. The model fails to reproduce the elementary data on the proton in the region between $\Delta$ and second resonance region, but it describes the nuclear data in a satisfactory manner. 
The work of Hirata~\cite{Hirata:2001sw} points out that according to their model the interference patterns among the resonances and the background change when going from vacuum to medium. In their model, this change of the interferences is the driving force for the disappearance of the resonances. 
Iljinov \etalEnde~\cite{Ilinov:1996js} extended the \textit{Dubna/Moscow INC} model, a hadronic transport model, such that it can be used for high photon energies up to $10 \GeV$. They achieved good correspondence with exclusive channels such as single-pion production. Based on the \textit{Dubna/Moscow INC} model, the \textit{RELDIS} code~\cite{Pshenichnov:2005zz} achieved good results for photo-absorption on large nuclei. Both the \textit{INC} and \textit{RELDIS} model include a phenomenological two-body absorption channel on top of single-particle absorption. Also the \textit{LAQGSM} model~\cite{Mashnik:2005dm} is based on the \textit{Dubna/Moscow INC} model, which also results in a good model for nuclear fissibilities. 
Deppman \etalEnde~\cite{Deppman:2002cg} have successfully applied the so-called \textit{MCMC/MCEF} cascade model, to evaluate photon fission cross sections using the photon absorption cross section as input. Complementary to this first work, Deppman achieved with the \textit{CRISP} code~\cite{Deppman:2004vc,Deppman:2006gb}, where the photon absorption is modeled via a microscopic resonance model, also satisfactory results for photon absorption.

Effenberger \etalEnde~\cite{Effenberger:1996im} tried to explain the experimental observation of \textit{resonance disappearance} within a precursor version of our present model. He achieved a quite good description of the $\Delta$ peak, however his model failed explaining the data at higher energies: his results showed still a prominent structure in the second and third resonance region. 
Our present model differs to Effenberger's one in four major aspects. First, we use the up-to-date input for the photon-nucleus interaction, in particular for the helicity amplitudes of the resonances, and for the invariant amplitudes used to parameterize pion production. Second, we treat the single-pion background terms differently: Effenberger \etal rescaled the resonance contributions to fit the single-pion data while we include a point-like single-pion production vertex which includes all background and interference terms. So in our case the background and interference terms are not modified in the medium, whereas in Effenberger's case those two terms are hidden in the resonance contributions and, therefore, they are modified in the same way as the resonances. Third, the absolute magnitude of the collisional width is slightly different due to updated baryon-nucleon cross sections in the GiBUU collision term. Finally, Effenberger only included the mean fields as real part of the self energies and neglected the dispersive contributions to the real parts, which lead to non-normalized spectral functions. 

\begin{figure}
\centering
\includegraphics[width=0.45\textwidth]{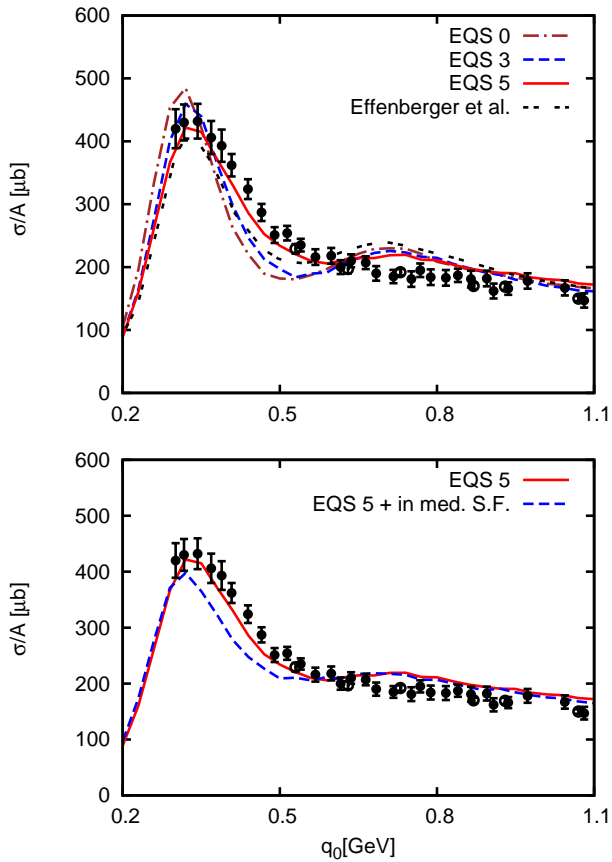} 
\caption{The photon absorption cross section for $^{12}C$. In the upper panel we show the the impact of different assumptions on the baryon potential: the red solid line denotes the result with a momentum-dependent potential (EQS 5), the dashed blue one represents a calculation with a momentum-independent potential (EQS 3) and the brown dashed-dotted a calculation where the baryon potentials have been neglected (EQS 0). For details on the potentials confer~\cite{Teis:1996kx}. In the lower panel we keep the potential the same (EQS 5) and add a modification of the in-medium self-energy. The data are taken from Bianchi \etal \cite{Bianchi:1995vb} (full circles) and Muccifora \etalEnde~\cite{Muccifora:1998ct} (open circles), the error bars denote the sum of statistical and systematical errors.} 
\label{fig:photo_abs_C}
\end{figure}
Our results for photon absorption in Carbon are shown in \figref{fig:photo_abs_C}. For the calculations shown in the upper panel, we have not modified the width of the resonances in the medium and we have not included dispersive contributions to the width. Again, as for the quasi-elastic peak in the electron-nucleus scattering, we observe the broadening of the spectra just due to the momentum-dependent mean field. The solid curve shows the result when including such a mean-field. This curve describes the data properly in the $\Delta$ region but overestimates it in the second resonance region. It is, however, remarkable, that the mean field smears the spectra so much that the second resonance peak is dissolved. 
The lower panel shows, additionally to the result calculated with a momentum-dependent mean field (solid line), the one for which the in-medium width and the dispersive contributions to the real parts have been taken into account (dashed line). A larger width due to the collisional broadening leads to a smearing of the peaks and therefore to a lowering of the resonance contributions. The lower panel shows that the $\Delta$ region is underestimated when including the in-medium spectral functions. Also in the second resonance region the situation does not improve. The same effect has also been observed in Effenberger's work where the missing strength in the right shoulder of the $\Delta$ is attributed to (also in our work) neglected two-nucleon absorption processes. Comparing to this former result by Effenberger (double dashed curves in the upper panels), we must state that both models describe the data equally well. In our approach it is however interesting to see, that the proper treatment of the medium-dependent potential leads all by itself to a smearing of the second resonance peak. As a consequence, it should be interesting to include the in-medium kinematics also in the background treatment, which was too involved for our approach since one would need a microscopic model for the background. 

\begin{figure}
\centering
\includegraphics[width=0.45\textwidth]{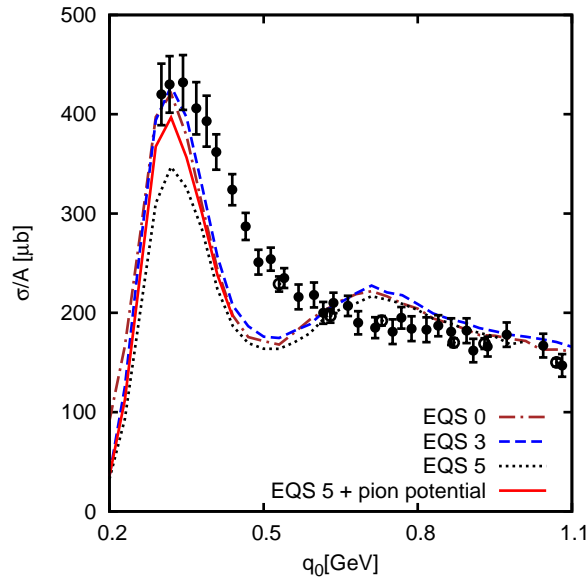} 
\caption{The photon absorption cross section for $^{12}C$ without resonances in the intermediate state. The curves in the upper panel represent calculation where we did not include resonances but treated the $\gamma N\to N\pi$ reaction as a point-like vertex. The red solid line denotes the result with a momentum-dependent potential (EQS 5) and a low-energy potential for the pion according to \cite{Buss:2006vh}, the other curves have been obtained without a pion potential using different types of the nucleon potential: no potential (dot-dashed), a momentum independent potential (dashed) and a momentum dependent potential (dotted curve). The data are taken from Bianchi \etal \cite{Bianchi:1995vb} (full circles) and Muccifora \etalEnde~\cite{Muccifora:1998ct} (open circles), the error bars denote the sum of statistical and systematical errors.} 
\label{fig:photo_abs_C_noRes}
\end{figure}

Fig.~\ref{fig:photo_abs_C_noRes} shows again the photon absorption in Carbon. To obtain the results shown in the latter graph, we have ignored the resonance model and have just included the point-like $\gamma N\to N\pi$ process, which we evaluated now with full in-medium kinematics \red{taking into account potentials for the nucleon and the pion. For the invariant amplitudes $A_i$ we assume the following in-medium structure 
\eb
A_i^\text{medium}(s,Q^2,\theta_k)=A_i^\text{vacuum}(s_\text{free},Q^2,\theta_k) \eKomma
\label{eq:ai_med}
\ee
with $s_\text{free}$ being defined in \eq{\ref{eq:sfree}} and $\theta_k$ is the pion scattering angle with respect to $\vec{q}$. As before we added also a $2\pi$ production channel, which is without resonance contributions a pure background contribution. In this approach, one does not account for in-medium changes of the width of the resonances and for potential differences of nucleons and resonances.} The $\Delta$ peak structure is in its magnitude well described by a calculation which includes both a pion potential according to \cite{Buss:2006vh} and a momentum-dependent nucleon potential (red solid line). However, the width of the $\Delta$ peak is underestimated and the peak structure in the second resonance region is not present in the data.  \red{Besides the missing in-medium changes of the resonances, also the approximation of \eq{\ref{eq:ai_med}} might be questionable. However, as long as we have no underlying model to evaluate the $A_i$ in the medium there seems to be no better choice. We conclude that the resonance model framework gives a far better description of the data and incorporates the in-medium changes in a more controlled fashion.}

%% file: summary.tex
We have presented a refined and updated treatment of electron scattering off nuclei. Employing in-medium kinematics and parametrizing the form factors for QE-scattering according to Bradford \etalEnde~\cite{Bradford:2006yz}, using the $\pi$-production form-factors and resonance helicity amplitudes of the MAID~\cite{MAIDWebsite,Drechsel:1992pn,Pasquini:2007fw} analysis. In comparison to the data by Anghinolfi \etalEnde~\cite{Anghinolfi:1996vm} we found good agreement in the pion production energy regime, whereas the quasi-elastic scattering is overestimated for low photon energies. We outlined the large dependency on the underlying momentum distribution of the nucleons, which we fix in our model via the LTF assumption. 

Additionally, we investigated photon absorption where we could not establish an improvement of earlier results. However, we emphasized the importance of momentum-dependent potentials for the process. In particular, the structure in the second and third resonance region gets dissolved due to this potentials. 

In future we plan to extent the description to exclusive channels, such as e.g. single $\pi$ production, and plan to include the $\pi\pi$ production channel also for electron scattering.